# Lung cancer screening with low-dose CT scans using a deep learning approach

**Jason L. Causey[1†], Yuanfang Guan[2†], Wei Dong[3], Karl Walker[4], Jake A. Qualls[1], Fred Prior[5*], Xiuzhen Huang[1*]**

[1]Department of Computer Science, Arkansas State University, Jonesboro, Arkansas 72467, United States of America

[2]Department of Computational Medicine & Bioinformatics, University of Michigan, Ann Arbor, Michigan 48109 United States of America

[3]Ann Arbor Algorithm, Ann Arbor, Michigan 48103, United States of America

[4]Department of Mathematics and Computer Science, University of Arkansas at Pine Bluff, Pine Bluff, Arkansas 55455, United States of America

[5]Department of Biomedical Informatics, University of Arkansas for Medical Sciences, Little Rock, Arkansas 72205, United States of America

† The authors are considered as joint first authors.
*Correspondence to: F. Prior (fwprior@uams.edu), X. Huang (xhuang@astate.edu)

**ABSTRACT**

Lung cancer is the leading cause of cancer deaths. Early detection through low-dose computed tomography (CT) screening has been shown to significantly reduce mortality but suffers from a high false positive rate that leads to unnecessary diagnostic procedures. Quantitative image analysis coupled to deep learning techniques has the potential to reduce this false positive rate. We conducted a computational analysis of 1449 low-dose CT studies drawn from the National Lung Screening Trial (NLST) cohort. We applied to this cohort our newly developed algorithm, DeepScreener, which is based on a novel deep learning approach. The algorithm, after the training process using about 3000 CT studies, does not require lung nodule annotations to conduct cancer prediction. The algorithm uses consecutive slices and multi-task features to determine whether a nodule is likely to be cancer, and a spatial pyramid to detect nodules at different scales. We find that the algorithm can predict a patient's cancer status from a volumetric lung CT image with high accuracy (78.2%, with area under the Receiver Operating Characteristic curve (AUC) of 0.858). Our preliminary framework ranked 16th of 1972 teams (top 1%) in the Data Science Bowl 2017 (DSB2017) competition, based on the challenge datasets. We report here the application of DeepScreener on an independent NLST test set. This study indicates that the deep learning approach has the potential to significantly reduce the false positive rate in lung cancer screening with low-dose CT scans.

**INTRODUCTION**

Lung cancer is the leading cause of cancer deaths and the second most common cancer in both men and women in the United States [1]. Because lung cancer is most often diagnosed at an advanced stage, the overall 5-year survival is poor (at 18%). Early detection is key to early intervention leading to improved survival. Compared with radiographs, low-dose CT can provide more detailed information and has been reported to lead to a 20% lower mortality rate [2]. Low-dose CT has been recommended for lung cancer screening by the US Preventive Services Task Force [3].

Lung cancer screening studies based on analysis by human experts have reported false positive rates as high as 58% [4]. This not only increases the cost for further tests and surgical procedures, but also causes unnecessary anxiety for patients and their families. The development of powerful computer-aided approaches for lung cancer early screening is critical to improve the current clinical practice of CT imaging assessment. Computer-aided approaches are expected to produce automated solutions for early lung cancer screening and a reduced false positive rate in diagnosis.

Numerous computer-aided approaches have been developed for chest image analysis in the past fifty years. Ginneken [5] reviews computer analysis in chest imaging and illustrates how the three types of approaches — rule-based image processing, machine learning, and deep learning — have been applied, and how deep learning is currently becoming the dominant approach with very promising results [6]. Most computational approaches to date such as [7-11], focus on finding and analyzing nodules in lung CT images. This dependency on predefined objects of interest requires detection and segmentation steps that are difficult to automate and limit the applicability of these approaches for automated screening.

We have developed a deep learning approach we call DeepScreener, based on a convolutional neural network (CNN) model. The algorithm uses a pseudo-3-D model considering context information of consecutive slices of a participant's lungs. DeepScreener is an automated end-to-end solution for lung cancer screening that does not require *a-priori* lung nodule annotations. The preliminary framework of the algorithm ranked 16th of 1972 teams (top 1%) in the DSB2017 competition [14]. We evaluated the algorithm's ability to generalize beyond the competition dataset by applying it to an independent cohort drawn from the National Lung Screening Trial (NLST) [2].

## RESULTS
**DeepScreener achieves high accuracy for lung cancer detection on the NLST cohort**.

We tested the performance of DeepScreener using 1449 low-dose CT studies obtained from the Cancer Imaging Archive [16] with permission from the National Cancer Institute. DeepScreener was able to make predictions for 1359 of 1449 CT scans with an accuracy of 78.2%, AUC of 0.858, the area under Precision-Recall curve (AUPRC) of 0.788, and log-loss of 0.484. Refer to **Figure 1** for the ROC curve and the Precision-Recall curve resulting from this analysis. The algorithm correctly identified 148 of 432 positive examples (sensitivity 34.3%) and 915 of 927 negative examples (specificity 98.7%). Refer to **Table 1** for the complete set of validation metrics and performance details, and the discussion section for our analysis of these results.

We found that several of the images in our test cohort exhibited some kind of unusual defect. For example, some images contained uneven "slice spacing" — the spacing between slices along the superior/inferior axis was not consistent. If the spacing varied enough to suggest a "gap", or missing slice, the image was rejected. We also noted that some images contained one or more "duplicate" slices — the pixels in the 2-D slice were identical to the pixels in another 2-D slice within the scan. In these cases, we dropped the duplicate slice in preprocessing. In total, DeepScreener rejected 90 CT studies due to inconsistencies.

**Evaluation Metrics.** We used the following statistical criteria to test the performance of DeepScreener: Accuracy, sensitivity, specificity, AUC, the f1-score and LogLoss. Accuracy is defined as $\text{acc} = \frac{\text{TP}+\text{TN}}{\text{N}}$ where $\text{TP}$ represents the number of true positives, $\text{TN}$ represents the number of true negatives, and $\text{N}$ represents the total number of scans considered. Sensitivity is defined as $\text{sen} = \frac{\text{TP}}{\text{TP}+\text{FN}}$ where $\text{TP}$ is the number of true positives and $\text{FN}$ is the total number of false negative (i.e. missed positive) scans. Specificity is defined as $\text{spc} = \frac{\text{TN}}{\text{TN}+\text{FP}}$ where $\text{TN}$ is the number of true negatives and $\text{FP}$ is the number of false-positive scans. We use AUC to refer to the area under the receiver operating characteristic curve, which plots the true positive rate against the false positive rate under varying classification threshold values [22]. We use AUPRC to refer to the area under the Precision-Recall curve, which plots the trade-off between precision and recall (recall is synonymous with sensitivity). The f1-score is a measure of accuracy involving both precision and sensitivity [23], defined as $F_1 = 2 \cdot \frac{\text{Precision} \cdot \text{Sensitivity}}{\text{Precision} + \text{Sensitivity}}$ where $\text{Precision} = \frac{\text{TP}}{\text{TP}+\text{FP}}$. LogLoss is defined as $\text{LogLoss} = -\frac{1}{\text{N}}\sum_{i=1}^{\text{N}}[y_i \cdot \log_e(\hat{y}_i) + (1-y_i) \cdot \log_e(1-\hat{y}_i)]$, where $\hat{y}_i$ is the predicted probability of the image belonging to a patient with cancer, $y_i$ is 1 if the diagnosis is cancer, 0 otherwise, $\log_e(\cdot)$ is the natural (base $e$) logarithm.

Table 1: Validation metrics for DeepScreener and grt123 algorithms on the NLST cohort.

| Metric | DeepScreener | grt123 |
| --- | --- | --- |
| Total | 1359 | 1449 |
| # Positive | 432 | 469 |
| # Negative | 927 | 980 |
| AUC | 0.858 | 0.885 |
| AUPRC | 0.788 | 0.837 |
| Accuracy | 0.782 | 0.821 |
| LogLoss | 0.484 | 0.434 |
| f1-score | 0.500 | 0.631 |
| Sensitivity | 0.343 | 0.473 |
| Specificity | 0.987 | 0.987 |
| # False Positives | 12 | 13 |
| # False Negatives | 284 | 247 |

**DeepScreener has performance comparable to the winning algorithm of DSB2017 on the NLST cohort**.

We compare the performance of our algorithm DeepScreener with the winning algorithm grt123 of Data Science Bowl 2017 for lung cancer detection. We find on this NLST cohort of 1449 low-dose CT scans, the performance of DeepScreener is very close to the performance of grt123. The algorithm, *grt123,* was able to process 1449 of the 1449 CT scans with accuracy of 82.1%, AUC of 0.885, area under Precision-Recall curve (AUPRC) of 0.837, and log-loss of 0.434. Refer to **Figure 2** for the ROC curve and the Precision-Recall curve. The algorithm correctly identified 222 of 469 positive examples (sensitivity 47.3%) and 967 of 980 negative examples (specificity 98.7%). Refer to **Table 1** for the performance details. DeepScreener achieved an AUC of 0.858, slightly lower than grt123. To determine whether the difference in performance was significant, we compared the Receiver Operating Characteristic curves for both methods using the DeLong method [24], with the null hypothesis that the difference in AUC for the two models is equal to zero. The test shows that with $p = 0.070$ we cannot reject our null hypothesis to a confidence level of $\alpha = 0.05$, thus we conclude that there is no significant difference in performance between the two methods on the NLST cohort.

**SUMMARY AND DISCUSSION**

Computed tomography screening has been shown to aid in early detection of lung cancer in at risk patients, leading to reductions in lung cancer death rates [2]. Unfortunately, CT screening is also associated with high rates of false-positive diagnoses. Currently most computer-aided diagnosis (CAD) tools focus on evaluating lung nodules, which must be identified *a-priori*, either by a radiologist or with an automated tool. Here we chose to instead focus on risk prediction at the patient level, taking into account information from the whole lung. Our approach could be combined with others to provide a layered strategy for identifying and diagnosing lung cancer. Our approach combines convolutional neural network models and an XGboost classifier to predict the presence of lung cancer at the whole-image level. We chose to test our strategy on low-dose CT scan data from the National Lung Cancer Screening Trial (NLST). To our knowledge, this is the first time a whole lung CNN-based classifier has been tested on this large NLST cohort.

On the NLST cohort of 1449 low-dose CT scans, we tested our deep learning algorithm for predicting lung cancer status with whole low-dose CT scans of the patients. Our algorithm DeepScreener was able to make predictions with high accuracy, with an AUC of 0.858, and an AUPRC of 0.788. From the testing results on the NLST cohort, we anticipate deep learning algorithms can achieve a performance potentially comparable to human experts and radiologists for lung cancer prediction and detection with low-dose CT scans. Through training and learning from CT images of even a larger population, the approach will yield sensitive, stable, consistent and reliable lung cancer screening with the potential of reducing the human effort and cost of screening.

The initial framework for DeepScreener was developed for the DSB2017 competition, and was optimized according to the performance metric used by the competition (minimizing log-loss) [14]. One of our goals for developing automated screening tools is to reduce the false-positive rate associated with lung screenings performed by radiologists. However, the particular metric chosen by the DSB2017 competition may have skewed the model too far in the direction of reducing false positives at the expense of missed cancers (false negative rate). In addition, the cohort we used for this validation procedure was selected to include cases where the original NLST study contained a likely "false negative" screen. We did this by querying for cases where the patient's cancer diagnosis followed a negative screen and adding those cases to our query for patients who screened as "positive". Thus, our testing cohort itself may be expected to elicit a higher-than-normal false negative rate. More work needs be done to balance this trade-off to levels that are clinically acceptable. In evaluating both our competition model and the winning model against a previously-unseen set of challenging CT screening images, we gain some insight into the value of such competition models in real-world applications. Our results hint that the choice of a competition scoring metric may induce biases in the models that need to be addressed before wider application. We note that our model had an accuracy (78.2%) and an AUC (0.858). We examined the effect of modifying the decision threshold (which was at the default 0.5 for the results reported here) with respect to the model's output predictions on NLST and found that a lower threshold would have improved the results. A threshold setting of 0.29 would maximize the accuracy metric at 83.4%, and a threshold setting of 0.19 would maximize sum of sensitivity and specificity at 69.2% and 87.6% respectively. Obviously, any such tuning based on *a-posteriori* observations would need to be evaluated by re-training the model and performing an independent validation, but it suggests that training with an objective different than log-loss may be beneficial, as well as parameter tuning to guide the algorithm toward more balanced performance. In the future, we hope to further develop the model to decrease the number of missed positives, and also add visualization options to help make the model's classification decisions interpretable for researchers and clinicians. These improvements will be necessary before tools such as these will be accepted into the clinical diagnostic tool set.

**MATERIAL AND METHOD**

**Datasets for training, validation, and testing.** We use the following datasets: (1) the LIDC/IDRI cohort data, (2) the LUNA16 Challenge data, (3) the DSB2017 Competition data, and (4) a subset of the NLST cohort data. Note that there are some overlaps between these datasets, and we have carefully considered this and tried to remove the overlaps between the training and evaluation datasets.

The Lung Image Database Consortium image collection (LIDC/IDRI) [12] consists of diagnostic CT data sets with annotated lesions for 1018 participants. Each study includes images from a clinical thoracic CT scan and an associated XML file that records the results of a two-phase image annotation process performed by four experienced thoracic radiologists. The radiologists annotated each scan by marking regions of interest in three classes: "nodule $\geq$ 3mm", "nodule < 3mm", and "non-nodule". Each nodule in the "nodule $\geq$ 3mm" class was then given a malignancy score and a detailed segmentation. The LUNA16 Challenge [13] released a list of additional nodules, which were missed by expert readers who originally annotated the LIDC/IDRI data.

The Kaggle Data Science Bowl 2017 [14] dataset is comprised of 2101 chest CT studies. Of the 2101, 1595 were initially released in stage 1 of the challenge, with 1397 belonging to the training set and 198 belonging to the testing set. The remaining 506 were released in stage 2 as a final testing set. Each CT study was labeled as 'with cancer' if the associated patient was diagnosed with cancer within one year of the scan, and 'without cancer' otherwise. Crucially, the location or size of nodules are not labeled. This data was partially drawn from the NLST cohort. Care was taken in selecting our test cohort to be independent, as explained below.

We tested the performance of the algorithm using data from National Lung Screening Trial (NLST) (with permission of NCI). 1663 screens were selected for this study with 1000 negative screens and 663 positive screens. The ground truth labels for each study were defined as the presence or absence of a cancer diagnosis during the NLST trial period [2]. We eliminated 5 screens due to missing image data at the time point where the screen was marked as "positive". An additional 209 screens were eliminated due to an overlap with training data from DSB2017 (202), LIDC/IDRI (3), or both (4).

To identify overlapping images, we used an image fingerprinting method based on comparing intensity histograms of selected slices from each scan in all three primary source cohorts (LIDC/IDRI, DSB2017, NLST). Note that LUNA16 is a subset of LIDC/IDRI. Fingerprints for individual slices were produced by loading the pixel values from the DICOM image representing the slice and transforming the pixel intensities to Hounsfield Units (HU). Then an intensity histogram containing 20 bins roughly centered in regions representing different tissue densities was generated. Bin boundaries were fixed to the following HU values: $[-1024, -500, -300, -150, -125, -100, -80, -40, -2, 0, 20, 40, 60, 80, 100, 125, 150, 300, 500, 1024, 2048]$. Histograms were calculated in this way for each of the first and last ten slices of each scan, ordered by the *Instance Number* DICOM attribute. The histograms were combined into a fingerprint vector that was utilized for comparing the mean squared error (MSE) of all possible combinations of images from each dataset. This method was chosen to be both relatively computationally efficient and robust against possible changes made to images when migrating from their original datasets into the competition cohort (such as re-sampling voxel dimensions, reversing the superior/inferior axis, or missing/duplicated slices). We found no evidence of such changes; all overlaps we discovered had an MSE < 0.001, with a large gap between matches and non-matches (MSE > 200). The lowest scoring non-matches were examined visually to confirm that they were not modified versions of the same scans.

**DeepScreener: a novel algorithm to predict lung cancer.** DeepScreener provides an automated solution to predict whether a patient has lung cancer based on a low-dose screening CT study. Please refer to **Figure 3** for an overview of the deep learning algorithm.

Our training strategy can be summarized as follows. First, we trained the image analysis stage of the model for nodule identification using the images and radiologist annotations for nodules in the LIDC/IDRI cohort. We also included annotations for additional nodules released by the LUNA16 Challenge. Then we trained the classification stage of the model for predicting lung cancer from CT data without *a-priori* nodule annotations. For this purpose, we used the DSB2017 Competition [14] stage one CT data for training, and the stage two CT data for validation. The classification stage was trained by minimizing log-loss with respect to the ground truth classifications provided by the DSB2017 competition. Finally, generalization testing of the algorithm was conducted with our own selected cohort of low-dose CT images from the NLST study, and reported here.

In the following we describe the algorithm DeepScreener in detail, including how to train the CNN models and make predictions to identify patients with lung cancer using low-dose whole CT scans.

**Pseudo 3-D Model to Extract Consecutive Information across Slices.** We view a patient CT scan as a 3-dimensional volume. For example, a typical chest CT scan is about 512 x 512 x N, where N is the number of slices. The resolution within the 2-D slices may be different than between slices. A standard convolutional network can only handle **2-D** data, and the scan has to be processed as N individual slices of 512 x 512. Such slice-based processing loses almost all contextual information along the third dimension. For example, a blood vessel facing the *z*-dimension (perpendicular to the image in an axial view orientation) appears as a sphere and might be mistaken for a small nodule. Note that a 3-D convolutional network could be used to handle the 3-D information, but a 3-D network has limitations. For example, compared with a 2-D model, a 3-D convolutional network has many more parameters and is therefore more difficult to train. Training a 3-D network typically requires a much larger training data set. Instead, we chose to use a pseudo-3-D model. Our approach takes advantage of the fact that an image can have multiple (typically 3) channels and encode neighboring slices as multiple channels of a single image. Specifically, for each slice processed, we use the slice itself as the **"green"** channel of the image, and add one slice above as the **"blue"** channel and once slice below as the **"red"** channel, each at a distance of 4mm; see **Figure 4**.

**Multi-Task Learning for Feature Extraction for Cancer Classification of the Detected Nodules.** A segmentation network [15, 17] only produces a 2-dimensional shape for each nodule detected, and the shape boundary is typically blurry due to low decision confidence. It is possible to extract a few features, like area, average confidence and aspect ratio, but such features extracted solely based on a 2-D shape cannot capture all characteristics of a nodule that are visible to an expert viewing the original volumetric image.

The LIDC/IDRI dataset [2] provides expert annotation of about 1000 CT scans. In addition to nodule contours, a series of descriptive features are provided for each nodule, e.g. subtlety, sphericity, lobulation, etc. We designed a multi-task convolutional network to simultaneously fit 9 such features (see **Figure 5**): subtlety, sphericity, margin, lobulation, spiculation, texture, malignancy, calcification-1 and calcification -2. We do not use all of the available categorical features provided by LIDC/IDRI because we found some to be redundant. We split the categorical feature "calcification" into two binary features. This feature extraction network can increase the information available to the subsequent machine-learning module, i.e. gradient boosting decision trees (GBDT) [18] and improve classification stage accuracy.

**Feature Pooling with Spatial Pyramid to Detect Tumors of Different Scales.** After segmentation, nodule detection and feature extraction, we convert each CT volume into a list of nodule location (x, y, z) and features (size, subtlety, etc). For subsequent learning with GBDT, we need to pool this list of variable length into a vector of fixed number of dimensions. We apply the spatial pyramid approach [19] for such pooling. Specifically, we define a fixed number of regions with overlap, by partitioning the 3-D volume in multiple ways. The image in Figure 6 shows two sample partitions, each with four regions. For each region, we use the feature vector of the largest nodule as the region feature vector, or zeros if no nodule is detected within this region. We then concatenate the feature vectors from all regions to produce a feature vector representing the full CT volume. Even though the spatial pyramid generates a holistic representation of a full CT scan, we can also use it to represent an individual nodule, simply by removing all other nodules detected from the same CT scan. In this way we can apply the GBDT classifier model to assign a confidence score for each individual nodule. The patient-level classifier utilizes this ensemble of scores to produce a single confidence score.

**ACKNOWLEDGMENTS AND FUNDING**

This work was partially supported by National Institute of Health NCI grant U01CA187013, and National Science Foundation with grant number 1452211, 1553680, and 1723529, National Institute of Health grant R01LM012601, as well as partially supported by National Institute of Health grant from the National Institute of General Medical Sciences (P20GM103429).

**AUTHOR CONTRIBUTIONS**

XH and FP oversaw the project. XH, JC, YG, and FP conceived and designed the study. JC, YG, and WD performed the experiments, worked on numerical testing. XH, FP, JC, YG, WD, KW, and JQ analyzed the data, evaluated the methods. XH, JC, and FP wrote the paper, with input from all authors.

**COMPETING INTERESTS**

The authors have declared that no competing interests exist.

**FIGURES**

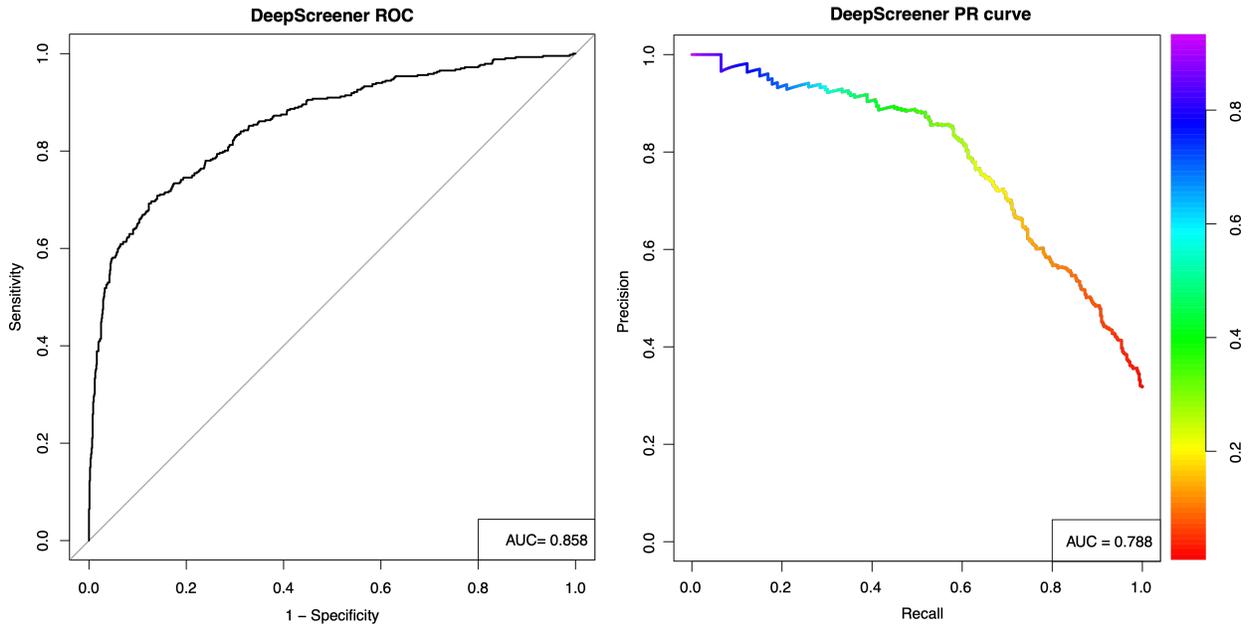

**Figure 1:** Performance characteristics for the DeepScreener algorithm applied to the selected NLST subset (N=1359). **(a)** Receiver operating characteristic curve. Area under the ROC curve is 0.858. **(b)** Precision-Recall curve. Area under the PR curve is 0.788.

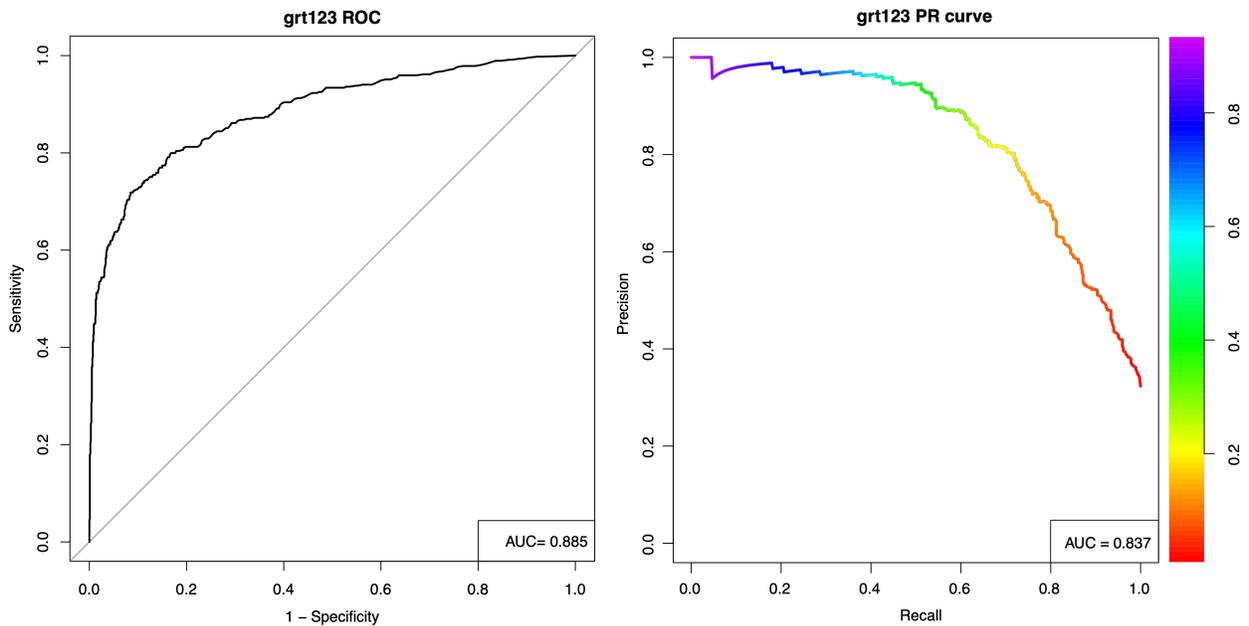

**Figure 2:** Performance characteristics for the *grt123* algorithm applied to the selected NLST subset (N=1449). **(a)** Receiver operating characteristic curve. Area under the ROC curve is 0.885. **(b)** Precision-Recall curve. Area under the PR curve is 0.837.

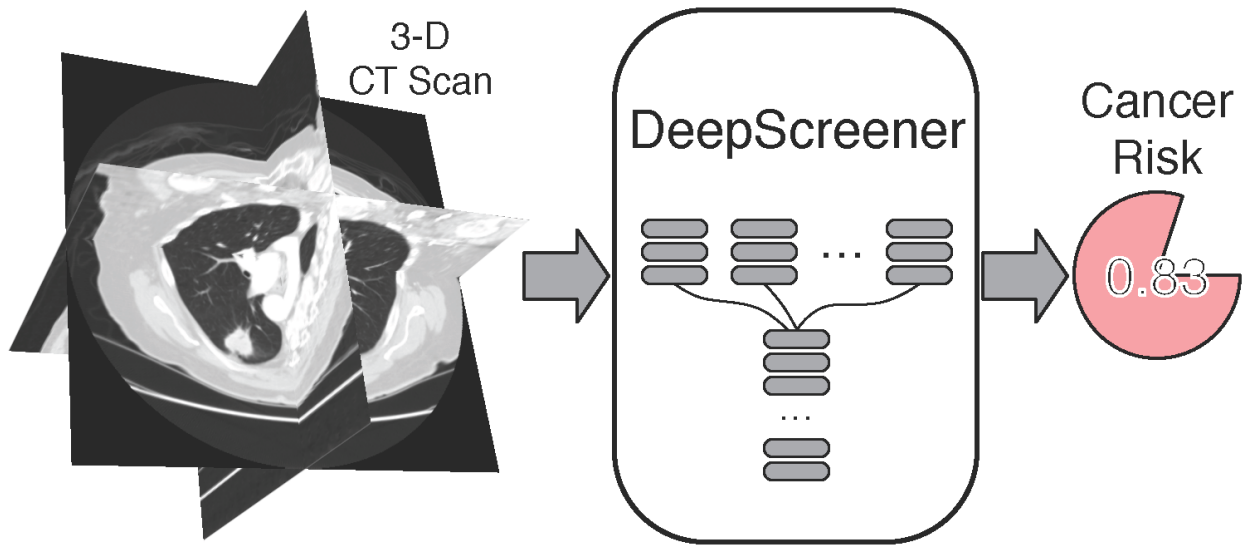

**Figure 3:** DeepScreener takes a 3-D chest CT image as input and uses a classification model based on convolutional neural networks and gradient boosting decision trees to output a prediction (in the range [0,1]) representing the likelihood that the patient has lung cancer.

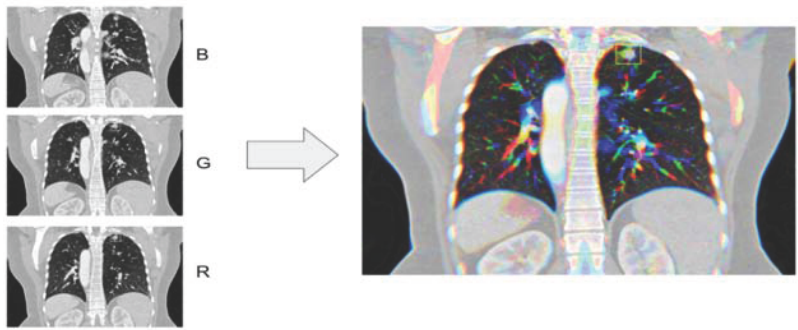

**Figure 4:** Pseudo-3-D image produced by "stacking" three individual slices from the 3-D CT image into the blue (B), green (G), and red (R) color channels of a 2-D 3-channel image.

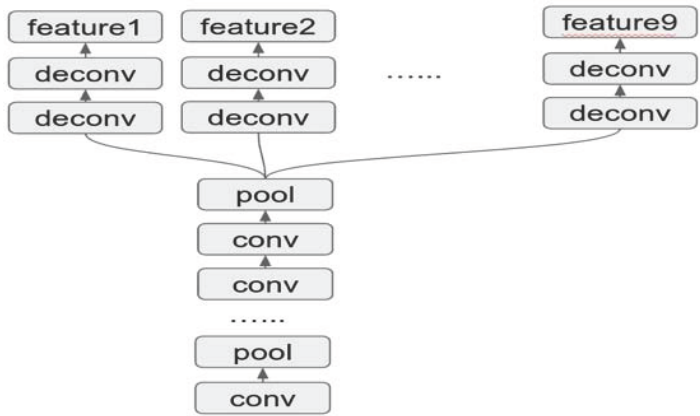

**Figure 5:** Multi-task convolutional network simultaneously computes 9 features as annotated in the LIDC/IDRI cohort: subtlety, sphericity, margin, lobulation, spiculation, texture, malignancy, calcification-1 and calcification-2.

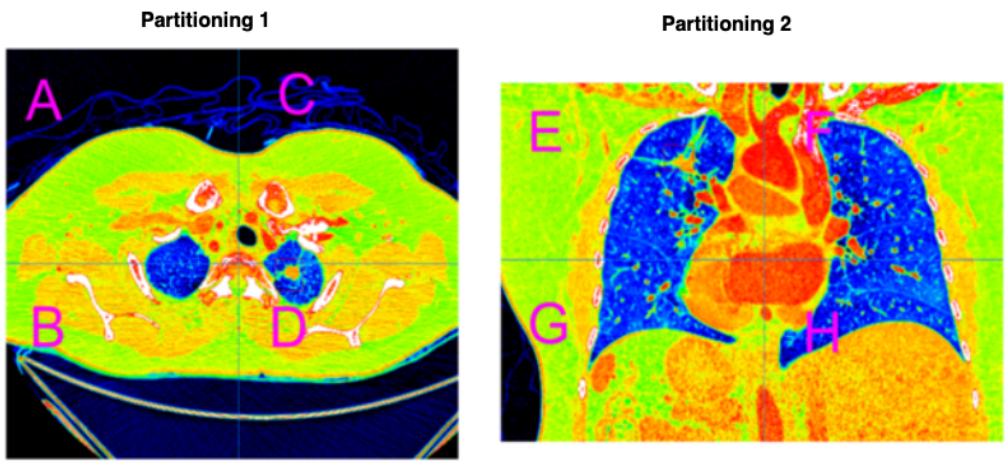

**Figure 6:** Example of two partitioning schemes: Each partitioning scheme defines four regions, which may overlap.